***Research Article***

# *The Vegetation Red Edge Biosignature Through Time on Earth and Exoplanets*

Jack T. O'Malley-James[1], Lisa Kaltenegger[1]

*1. Carl Sagan Institute, Cornell University, Ithaca, NY 14853, USA*

## Abstract

The high reflection of land vegetation in the near-infrared, the vegetation red edge (VRE), is often cited as a spectral biosignature for surface vegetation on exoplanets. The VRE is only a few percent change in reflectivity for a disk-integrated observation of present-day Earth. Here we show that the strength of Earth's VRE has increased over the past ~500 million years of land plant evolution and may continue to increase as solar luminosity increases and the planet warms, until either vegetation coverage is reduced, or the planet's atmosphere becomes opaque to light reflected off the surface. Early plants like mosses and liverworts, which dominated on land 500-400 million years ago, produce a weaker VRE, approximately half as strong as that of modern vegetation. We explore how the changes in land plants, as well as geological changes like ice coverage during ice-ages and interglacial periods, influence the detectability of the VRE through Earth's geological past. Our results show that the VRE has varied through the evolutionary history of land plants on Earth, and could continue to change into the future if hotter climate conditions became dominant, encouraging the spread of vegetation. Our findings suggest that older and hotter Earth-like planets are good targets for the search for a VRE signature. In addition, hot exoplanets and dry exoplanets with some water could be the best targets for a successful vegetation biosignature detection. As well as a strong red edge, lower cloud-fractions and low levels of atmospheric water vapor on such planets could make it easier to detect surface features in general.

## *1. Introduction*

Among several thousand detected exoplanets that now provide a first glimpse of the diversity of other worlds (e.g., reviewed in Udry & Santos 2007; Winn & Fabrycky 2015) are a few dozen small exoplanets that could potentially be habitable (see, e.g., Batalha 2014; Kane et al. 2016). Our neighboring star, Proxima Centauri, a cool M5V dwarf only 1.3 parsecs from the Sun, harbors a planet in its habitable zone (HZ) with a minimum mass of 1.3 Earth masses that receives about 65% of Earth's solar flux (Anglada-Escudé et al. 2016). At 3.4 parsecs (or 11 light-years) from the Sun, the planet Ross 128b, with a minimum mass of about 1.4 Earth masses, orbits in the HZ of its cool M4V dwarf star (Bonfils et al. 2017). The close-by TRAPPIST-1 planetary system of seven transiting Earth-sized planets around a cool M9V dwarf star, has several (three to four) Earth-sized planets in its HZ only about 12 parsecs from the Sun (e.g. Gillon et al. 2017; Ramirez & Kaltenegger 2017; O'Malley-James & Kaltenegger 2017). The planet LHS 1140b orbiting in the HZ of its cool M4V dwarf star, has a rocky composition based on its radius of 1.4 Earth radii and mass of 6.7 Earth masses (Dittmann et al. 2017). These four planetary systems already show several interesting close-by targets of potentially habitable worlds. A comprehensive suite of tools will be needed to characterize habitable planets and moons as the mere detection of a rocky body in the HZ does not guarantee that a planet is habitable (see e.g. review by Kaltenegger 2017). Signs of life that modify the atmosphere or the surface of a planet, and thus can be remotely detectable, are key to finding life on exoplanets or exomoons. Observations of our Earth with its diverse biota function as a Rosetta Stone to identify habitats.

Remote direct detection of surface life in reflected light from exoplanets becomes possible when organisms modify the detectable reflectivity of the surface (e.g. by influencing surface colors). Land vegetation is commonly cited as such a surface biosignature. On Earth vegetation has a specific reflection spectrum, with a sharp edge around 700 nm, called the Vegetation Red Edge, VRE.

Photosynthetic plants efficiently absorb visible light, but developed strong infrared reflectivity resulting in the steep change in reflectivity that defines the VRE. The advantages gained by reflecting this radiation are still debated. Some studies have claimed that plants balance the absorption of sunlight for photosynthesis reactions with efficient reflectance at other wavelengths to avoid damage caused by overheating (e.g. Gates et al. 1965; Seager 2005), but there is not yet enough data to determine consistent trends (Kiang et al. 2007). The primary molecules that absorb energy from light and convert it to drive photosynthesis ($H_2O$ and $CO_2$ into sugars and $O_2$) are chlorophyll a (absorption peaks at 430 nm and 662 nm in diethyl ether) and b (absorption peaks at 453 nm and 642 nm in diethyl ether) (Seager et al. 2005). The VRE is caused by the structure of leaves, which makes them very efficient at scattering light. Leaves are composed of water-filled cells surrounded by air gaps. Some light reflects off the cell walls; some enters the cells. The structure of leaf cells causes light entering them to be internally scattered (a combined effect of the high index of refraction between water and air, and the size of internal components of the cells, which are of the order of the wavelength of light). The scattered light eventually exits as transmitted light (below leaf) or reflected light (above leaf). Similar reflectance properties are found in other photosynthetic organisms, such as algae, which also exhibit red edges. This process enables the light-harvesting cells to absorb the specific wavelengths they need while effectively scattering away the rest. Chlorophyll pigments are almost transparent at wavelengths >700 nm, which results in the majority of the reflected/transmitted light being in the near-infrared (NIR) (e.g. Gates et al. 1965; Kiang et al. 2007, Seager et al. 2005 and references therein). The exact wavelength and strength of the spectroscopic VRE depends on the plant species and environment (see e.g. Kiang et al. 2007 and discussion in Rothschild 2008).

The VRE is often suggested as a direct signature of life (e.g. Sagan et al. 1993; Seager et al. 2005) and is commonly modeled using deciduous vegetation. The concept of a VRE biosignature is based on the assumptions that (i) extraterrestrial photosynthesizers evolve on other worlds, and (ii) these organisms also reflect strongly in the NIR. The first assumption is plausible for a wide range of stars – a planet's host star would be the most readily available source of energy for life and strong infrared reflectance would be a favorable method for protecting light-harvesting pigments (e.g. Wolstencroft & Raven 2002; Seager et al. 2005). The second assumption is plausible for planets around stars that have enough high-energy radiation to provide sufficient energy for photosynthesis, like Solar twins, but becomes less plausible for cooler stars (e.g. Kiang et al. 2007). A thorough analysis of the likelihood of oxygenic photosynthesis arising elsewhere is given by Wolstencroft & Raven (2002) and Rothschild (2008). As is the case for other biosignatures, we need to investigate how this signature has changed on Earth through time to gain knowledge of a wider range of possible strengths the signature could have on other worlds.

The VRE signature for the present day Earth is weak and very difficult to detect. Several groups (Arnold 2008 and references therein) have measured the integrated Earth spectrum via the technique of Earthshine, using sunlight reflected from the non-illuminated side of the moon. Averaged over a spatially unresolved hemisphere of Earth, the additional reflectivity of this spectral feature is typically only a few percent (e.g. Montañés-Rodriges et al. 2006) for present-day land plants (predominately deciduous vegetation).

Here, we explore how the VRE has changed through the evolution of land plants on Earth and project into its future. We use the reflectance spectra of different vegetation types to assess how the strength of Earth's red edge signature would change if different types of vegetation were dominant and how Earth's (or an Earth-like planet's) spectrum would change as a result of a change in dominant vegetation. We show disk-integrated spectra similar to the observation geometry obtained shortly before or after secondary eclipse or during direct imaging of the planet itself.

We also explore the range of appearances of these spectra using a standard astronomical tool, a color-color diagram, using standard Johnson-Cousins BVI broadband filters to define the color bands. Finally, we assess the implications this would have for the detectability of vegetation reflectance on exoplanets.

## *2. Methods*

We model planetary spectra with different vegetation surface coverage with a present-day Earth atmosphere using EXO-Prime (Kaltenegger & Sasselov 2009); a coupled 1D radiative-convective atmosphere code developed for rocky exoplanets, which models an Earth-like exoplanet’s atmosphere, its spectrum (see e.g. Kaltenegger et al. 2010; Rugheimer et al. 2013; 2015) as well as the UV environment on its surface (see Rugheimer et al. 2015; O’Malley-James & Kaltenegger 2017).

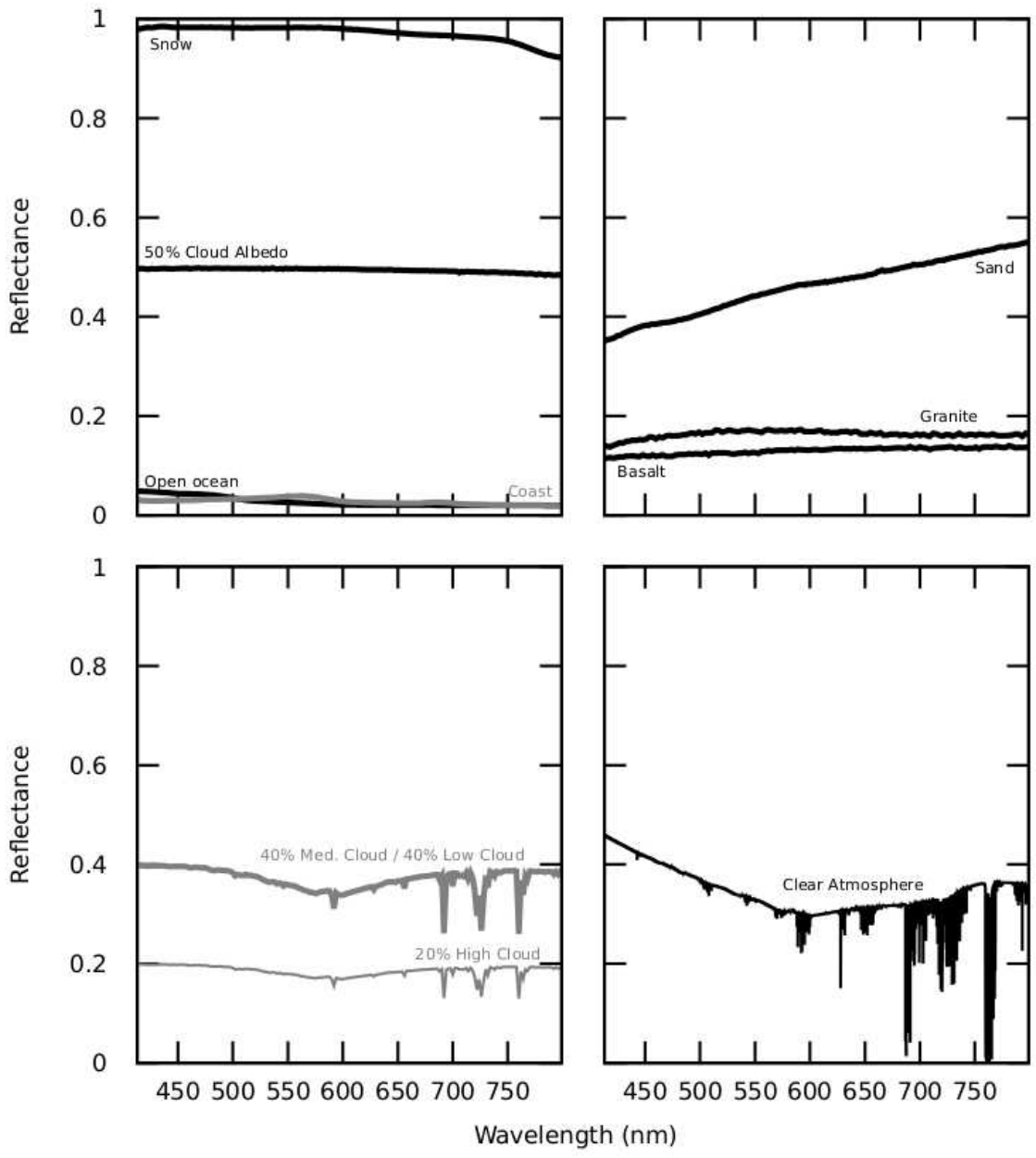

**FIG. 1.** Albedo and component spectra that make up our model Earth. The top panels show albedos of surface and cloud components and the lower panels show clouds and atmosphere spectra. The cloud profiles show the fractions of low-, medium- and high-altitude clouds on Earth (following Kaltenegger et al. 2007). Reflection spectra are from the USGS Spectral Library and ASTER Spectral Library.

Our radiative transfer model is based on a model that was originally developed to model the Earth's atmospheric spectra (Traub & Stier 1976) and has since been used extensively for analyzing high-resolution Fourier transform spectra from ongoing stratospheric balloon-based observations to study the photochemistry and transport of the Earth's atmosphere (for example, Jucks et al. 1998). Our line-by-line radiative transfer code has also been used for numerous full planetary disk modeling studies, both for theoretical studies (e.g., Des Marais et al. 2002; Kaltenegger et al 2010; Rugheimer et al 2015) and fitting observed earthshine spectra (e.g., Woolf et al. 2002; Turnbull et al. 2006; Kaltenegger et al 2007; Rugheimer et al. 2013). We divide the atmosphere into 60 thin layers from 0 to 100 km in altitude. The spectrum is calculated at very high spectral resolution, with several points per line width, where the line shapes and widths are computed using Doppler and pressure broadening on a line-by-line basis for each layer in the model atmosphere. We use a simple geometrical model in which the spherical Earth is modeled with a plane-parallel atmosphere and a single angle of incidence and reflection (visible) or emission (thermal infrared). This angle is selected to give the best analytical approximation to the integrated-Earth air mass factor of 2 for a nominal illumination (quadrature); the zenith angle of this ray is 60 degrees. The overall high-resolution spectrum is calculated at a resolution of 0.1 wavenumbers and smeared to lower resolution. For reference and further explanation concerning the code, the reader is referred to our calculation of a complete set of molecular constituent spectra, for a wide range of mixing ratios, for the present-day Earth pressure-temperature profile, and for the visible to thermal infrared, in Des Marais et al. (2002) and Kaltenegger et al. (2007).

We assume that the light paths through the atmosphere can be approximated by four parallel streams. All streams traverse the same molecular atmosphere, but each stream reflects (visible) or emits (thermal infrared) from a different lower surface. For example, in the visible and NIR, the first stream reflects from the planet's surface at 0 km altitude, the second and third streams reflect from a cloud layer with a top at an adjustable height (here 1 and 6 km for present-day Earth), and the fourth stream reflects from a cloud layer at a high altitude (here 12 km for present-day Earth).

To explore how different vegetation types affect the disk integrated spectrum of a planet and the VRE signal, we initially model a scenario where vegetation covers the entire the surface of the planet with an Earth composition atmosphere, modeling both clear-sky and 50% cloud-fraction scenarios. This initial model is then refined to represent the surface types on present-day Earth (see Fig. 1). We assign 70% of the planetary surface as ocean, 2% as coast, and 28% as land. The land surface consists of 60% vegetation, 9% granite, 9% basalt, 15% snow, and 7% sand for present-day Earth (following Kaltenegger et al. 2007). We also model changes to surface feature fractions during selected eras in Earth's geological history, as outlined in Table 2.

### *2.1. Color-color diagrams as a diagnostic tool*

We use a standard astronomy tool to characterize stellar objects, a color-color diagram, to distinguish the colors of planets with different surface spectra (following Hegde & Kaltenegger (2013)). To determine the difference between the reflectance, r, of two different color bands, we use equation (1):

$$C_{AB} = A - B = -2.5\log_{10}(r_A / r_B) \qquad (1)$$

where $C_{AB}$ is the difference between two arbitrary color bands, A and B. We use standard Johnson-Cousins BVI broadband filters to define the color bands (0.4 μm < B < 0.5 μm; 0.5 μm < V < 0.7 μm; 0.7 μm < I < 0.9 μm).

| Geological Period | Ocean % | Land % | Land Surface Vegetation % | Modelled Dominant vegetation type |
|---|---|---|---|---|
| Early Ordovician *475 Mya* | 81[a] | 19[a] | 10[d] | Moss |
| Ice Age (Ordovician glaciation) *440 Mya* | 81[a] (6.6% sea ice)[b] | 19[a] | 8.5[e] | Moss |
| Carboniferous *360-300 Mya* | 80[a] | 20[a] | 89.8[a] | Moss |
| Ice Age (late Carboniferous glaciation) *300 Mya* | 79[a] (11% sea ice)[c] | 21[a] | 45[e] | Ferns |
| Late Cretaceous *65 Mya* | 74[a] | 26[a] | 88.9[a] | Modern Earth |
| Ice Age (Last Glacial Maximum) *24.5 Kya* | 70[a] (11% sea ice)[c] | 30[a] | 45[e] | Modern Earth |
| Present-day Earth | 70[f] | 30[f] | 60[f] | Modern Earth |

**Table 1.** Description of surface feature distributions during the geological ages modeled and the type of dominant vegetation we assume in our models. *Sources:* **a.** Méndez et al. (2013) [Visible Paleo-Earth project]. **b.** Derived from sea ice area estimates from Poussart et al. (1999). **c.** Derived from sea ice area estimates from Otto-Bliesner et al. (2006) and assuming the late Carboniferous glaciation and Last Glacial Maximum had comparable ice extents (Barham et al. 2012). **d.** Based on estimates for vegetation land coverage following the first emergence of land plants from Lenton et al. (2012). **e.** For ice age vegetation fractions we subtract the increased land ice fraction (scaled from our value for the modern Earth using estimates from Méndez et al. 2013) from the vegetation fraction during the preceding period. **f.** Kaltenegger et al. (2007).

### *2.2. Model for Earth's Geological Surface Features Through Geological Time*

Rock records show that Earth's land, ocean and ice fractions have changed through geological time as climate and continental distributions changed. Therefore, the surface vegetation fraction will also have changed through geological time, but no clear consensus has yet been reached on paleo-vegetation distributions. Arnold et al. (2009) modeled vegetation distribution changes and the strength of Earth's VRE for two recent past climate extremes: the Last Glacial Maximum (4°C colder than present with large ice sheets) and the Holocene optimum (0.5°C warmer than present). In these cases, the strength of the VRE remained close to that of the modern Earth, with warmer temperatures inducing the greening of the Sahara, increasing the strength of the signal slightly (by ~1%). However, the impact of more extreme past climates from the last 500 Myr on the extent of planet land coverage and changes in continental configurations are still unknown (Arnold et al. 2009). Hence, we choose different points in Earth's history where surface fractions have been better constrained and use these to more realistically model the changes in the strength of the VRE from the emergence of land plants to the present day for our own planet (see Table 1).

## *3. Results*

The nature of NIR reflectance varies for different plant species. The magnitude of the increase in reflectivity and the wavelength at which the increase occurs both vary for different vegetation types. Deciduous vegetation on Earth causes an increase in reflectance in the NIR from approximately 5% to 50% when observed alone. For non-vascular plants like mosses, this reflectance peak can be as low as 20%, whereas for desert plants like cacti, it can be >80% (see Fig. 2). Note however, when the effect is hemispherically averaged and overlying clouds and an atmosphere are taken into account, the observed change is reduced to the few percent (Fig. 3 to Fig. 6) as observed in present-day Earthshine spectra (see also Montañés-Rodriges et al. 2006).

### *3.1. Timeline for Model of Vegetation coverage on Earth*

When land vegetation first emerged between 725 and 500 Myr ago, plants such as mosses, dominated the planet's surface, thus the spectral feature of deciduous vegetation would not have been present then. Modern flowering plants and trees have only dominated the surface for the past ~130 Myr (see Fig. 2). Stemming from an algal ancestor, the first land plants were simple and lacked a vascular system with which to transport water. However these soon evolved into larger leafy plants. The dominant surface vegetation on Earth would have changed during this time, potentially altering the strength of the spectral red edge signature. Cacti are highlighted here as an evolutionarily young (emerging as recently as 90-30 Mya – Nyffeler 2002) plant family (Cactaceae), which evolved after the split between South America and Africa, possibly as a result of an increase in aridity following an uplifting event in the Andes (Edwards et al. 2005). Cacti reflect red-to-infrared radiation more strongly than deciduous vegetation, making them interesting samples for red edge studies. All appearance times (see Fig. 2) are approximate, based on the earliest known fossil evidence (see for example Magallón et al. (2013) and references therein). Studies of carbon isotope ratios suggest that microbial photosynthetic life may have been present on land from as long as 1.2 billion years ago (Horodyski & Knauth, 1994; Raven 1997; Knauth & Kennedy 2009) long before the first land plants appeared – see Discussion.

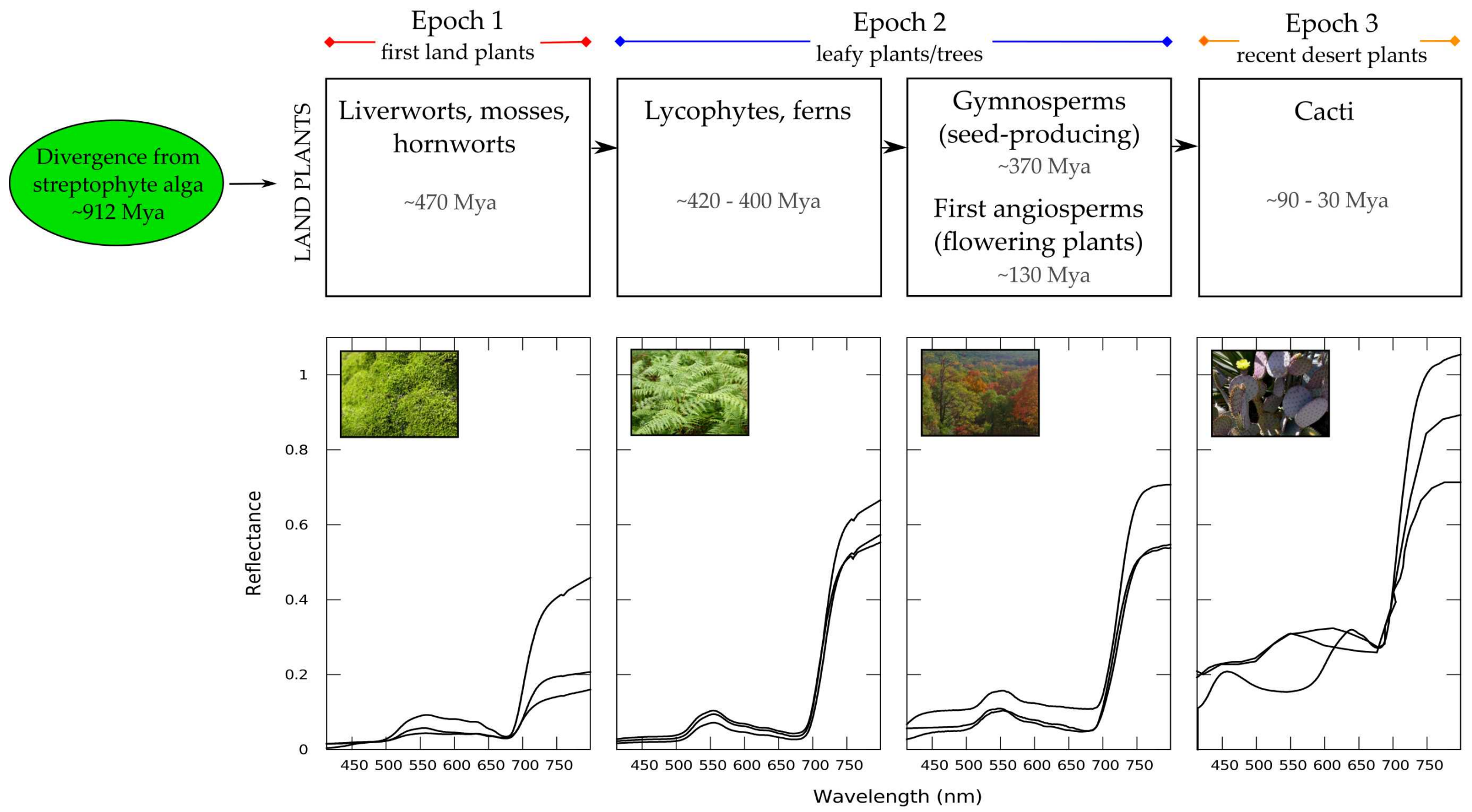

**FIG. 2.** (Top) Evolutionary time line for land plants. (Bottom) Reflectivity examples of different vegetation types for each of the stages outlined (moss, ferns, Gymnosperms, flowering planets and cacti). Deviations within each of group are generally small. Note that the reflective spines on some cacti can scatter light into the column of light in which reflectance is being measured, giving the appearance of >100% reflectance when observed alone. [Mya: Million years ago].

For this study we split the timeline into three epochs: Epoch 1, a young vegetation Earth (plants with low VRE strengths); Epoch 2, present-day Earth (plants with VRE strengths similar to present-day deciduous vegetation, from early ferns to modern trees); Epoch 3, a future Earth, (plants with high VRE strengths, which could represent a dominant vegetation type on a hotter Earth).

| *Species* | *Type* | *Source/Habitat* |
|---|---|---|
| **Moss sp. (unknown)[1]*** | **Non-vascular, mosses** | **Western Montana** |
| Moss sp. (unknown) [1]* | Non-vascular, mosses | Eastern Washington State |
| Liverwort sp. (unknown)[1]* | Non-vascular, liverworts | Interior Alaska |
| *Pteridium aquilinum[1]** | Vascular, ferns | Global (temperature, subtropical regions) |
| *Adiantum sp.[1]** | Vascular, ferns | Western Montana |
| *Adiantum sp.[1]** | Vascular, ferns | N. America |
| *Pinus edulis[2]*** | Vascular, gymnosperm (pine tree) | (SW) N. America |
| ***Elaeagnus angustifolia[2]**** | **Vascualar, angiosperm** | **Western/Central Asia** |
| Deciduous trees (mix)[3]** | Vascular, angiosperm | N. America |
| ***Opuntia violacea[2]*** | **Vascular, angiosperm (cactus)** | **N/S America** |
| *Opuntia gosseliniana[4]*** | Vascular, angiosperm (cactus) | N/S America |
| *Opuntia aciculat[4]*** | Vascular, angiosperm (cactus) | N/S America |

**Table 2.** Examples of the vegetation used that represent the range of red-edge strengths in the available spectral data. [1] Joint Fire Science Program spectral library, [2] USGS spectral library, [3] ASTER Spectral Library, [4] Gates et al. (1965). These are all single-leaf measurements with the exception of the "Deciduous trees (mix)", which uses leaf piles as a simulated canopy. Note that spectral deviations within each of group are generally small. Bold text highlights the species with the strongest red edge signals from each epoch (see Figure 2) that we selected to produce the plots in Figures 3-6.
**Data from in-situ measurements using hand-held field spectrometers. **Data from laboratory spectral measurements.*

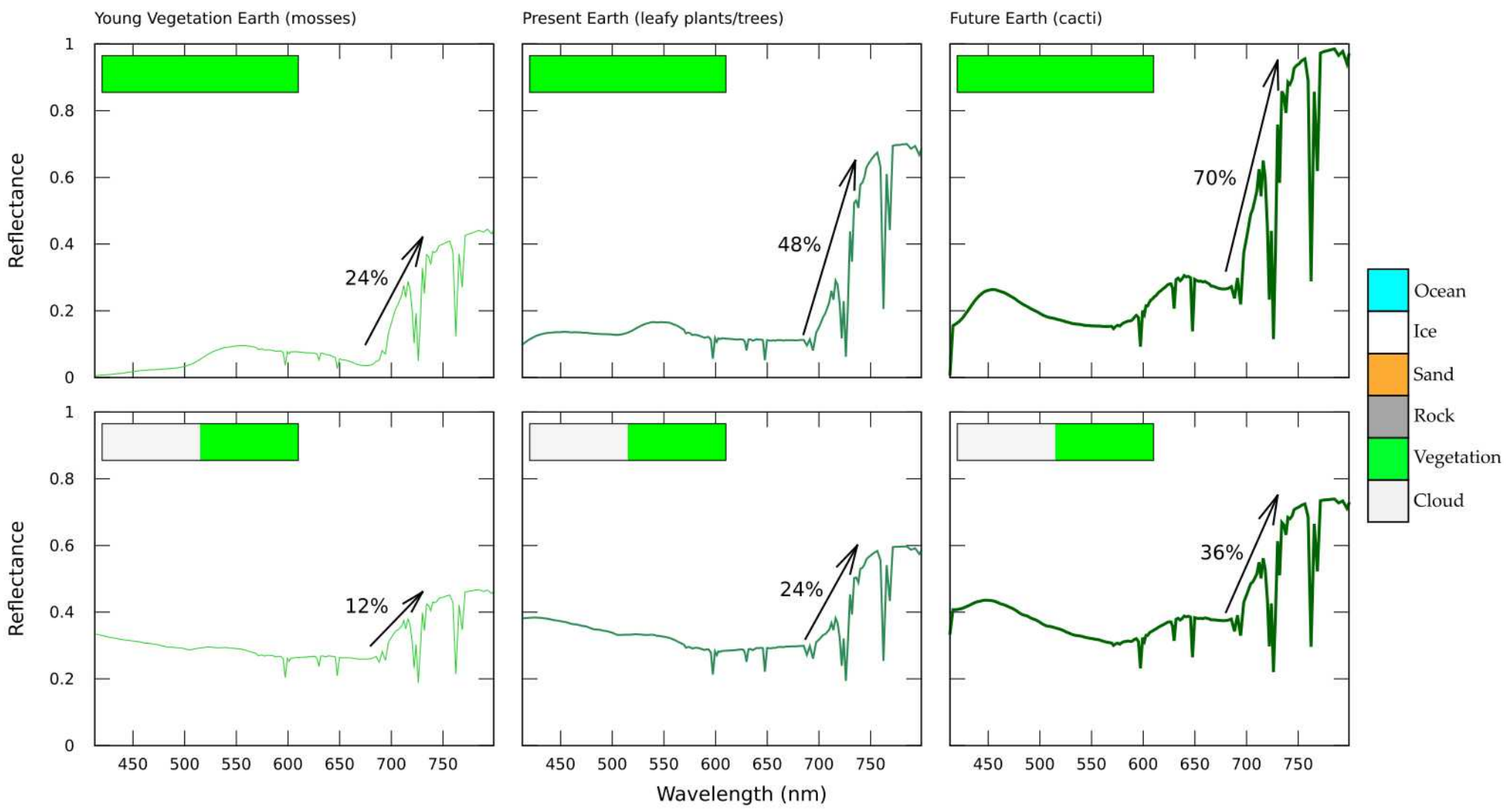

**FIG. 3.** Modeled VRE for 100% surface vegetation coverage for a "Young Vegetation" Earth ("Epoch 1"), present-day Earth ("Epoch 2") and a future hot, dry Earth ("Epoch 3") with a clear atmosphere (top) and 50% cloud cover (bottom).

Figure 2 illustrates the range of reflectivity of the land vegetation we explore (using the species listed in Table 2). All show a red edge feature, which varies in strength from species to species. As shown in Figure 2 the different mosses, ferns and tress have relatively little deviation within each group. Therefore we choose a species to represent the upper range of red edge strengths for each of the plant-type categories outlined in Fig. 2, which we identify in bold text in Table 2. We categorize these into three vegetation epochs representing different points in the evolutionary timeline of plants, which we match to seven eras in Earth's geological history (from 500 Mya to the present, as outlined in Table 1) and to the future-Earth. Mosses represent the main surface vegetation coverage for the earliest times when Earth hosted land vegetation, followed by ferns for later periods, deciduous vegetation for more recent periods and the present-day Earth. For a future hot, dry Earth we model cacti-dominated surface vegetation (see Discussion). Note that this timeline only represents the timeframe when land vegetation was widespread (starting < 500 Mya based on widely accepted fossil evidence, although land plants could potentially have emerged as early as 725 Mya; Magallón et al. (2013), Zimmer et al. 2007 (see Discussion)). Therefore we use a present-day Earth atmosphere composition for all our models.

### *3.2. VRE through geological time increases: Full Surface Coverage*

When a planet is fully covered by surface vegetation the VRE increases with geological time (Fig. 3) tracing the VRE strength increase in the evolution of the plants (see also Fig. 2). The VRE signal strength increases from 24% (Mosses) to 70% (cacti) for a clear atmosphere (see Fig.3 top). The signal strength decreases when we add clouds to the simulations because clouds are very reflective (Fig.1) and cover part of the surface from view. For an atmosphere with present-day Earth's 50% cloud fraction, the VRE strength increases through the evolutionary timeline of land plants from 12% (Mosses) to 35% (cacti); half of the values for a clear atmosphere.

### *3.3. VRE through geological time increases: Earth-like Ocean Coverage*

As a next step, we reduce the land cover to Earth-like conditions (28% continents, 2% coast and 70% oceans), however keeping the land surface vegetation covered. As shown in Figure 4, the overall VRE signal reduces due to the lower surface coverage. For a clear atmosphere the VRE for an Earth-like ocean coverage increases from 7% (Mosses) to 22% (cacti), for a planet with Earth's present-day cloud coverage of 50% the VRE increases from 3.5% (Mosses) to 11% (cacti), half of the values for a clear atmosphere.

### *3.4. VRE through geological time increases: Earth*

First we model the influence the changing land:ocean fraction over geological time has on the VRE, keeping the fraction of the non-frozen surface that is covered by vegetation constant throughout our selected time periods (Fig. 5). We look at the evolution of the VRE through geological time calling out warm epochs (Fig. 5A) and ice ages (Fig. 5B) separately, for a clear atmosphere (top) and an Earth-like cloud cover (bottom). Then we add information about Earth's evolution to our model, accounting for different surface vegetation fractions for all epochs we consider (see Table 1) showing the combined effect of changes in vegetation as well as surface properties for our own planet on the VRE signal.

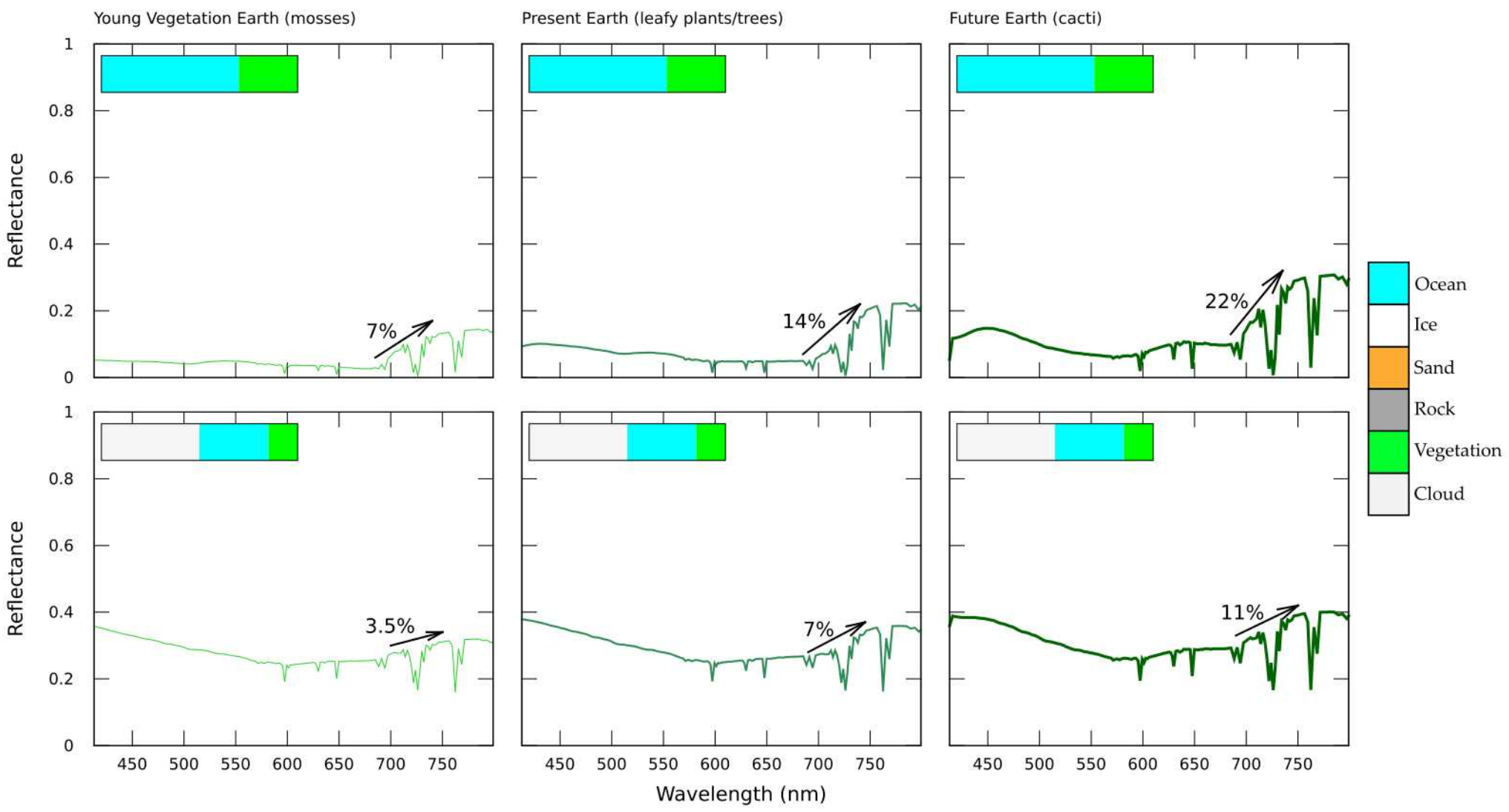

**FIG. 4.** Modeled VRE for present-day land:ocean fractions, with 100% vegetation on land, for a "Young Vegetation" Earth ("Epoch 1"), present-day Earth ("Epoch 2") and a future hot, dry Earth ("Epoch 3") with a clear atmosphere (top) and 50% cloud cover (bottom).

Figure 5 shows that for a clear atmosphere the VRE for warm epochs through Earth's history increases from 2.8% (Mosses) to 9% (present Earth). For present-day Earth's cloud coverage of 50% the VRE increases from 1.4% (Mosses) to 4.5% (present Earth). Present-day Earth shows a 4% increase in reflectivity due to vegetation in our models, in agreement with observations from Earth (see e.g. Arnold 2008; Robinson et al. 2014; Palle et al. 2016). A similar increasing trend is seen during cold epochs (Fig. 5B). For a clear atmosphere the VRE for Earth's cold epochs increases from 2.8% (Mosses) to 8% (modern Earth-like vegetation), for present-day Earth's cloud coverage the VRE increases from 1.4% (Mosses) to 4% (modern Earth-like vegetation).

The changes in land vegetation fractions over time (Fig. 6) modify the increase of the VRE over time. The increased reflectance caused by high surface ice fractions during ice ages, and corresponding decrease in surface vegetation cover, modulates the overall increase of the VRE signal. Note that hot, highly vegetated periods, like the Cretaceous, cause much higher (~10%) VRE strengths than on the present day Earth, making such epochs on an exoplanet very good targets for VRE detections.

### *3.5. Colors of Land Plant Evolution in a Color-Color Diagram through geological time are distinct from other solar system planets*

The differences between the colors of the different vegetation surfaces only are shown in Figure 7. Similar to the spectral results, the colors of a planet fully covered in each type of vegetation are clearly distinguishable (Fig. 7, top-left). Vegetation with a stronger red-edge, like cacti, tend to fall in the bottom-left quarter of the color-color diagram, while vegetation with weak red-edge features fall in the upper-right quarter of the diagram. Variations in visible pigmentation between species causes some deviation from this pattern.

Including cloud cover and/or an Earth-like land:ocean fraction pushes the points on the color-color diagram closer together, which helps to better identify the color space of vegetation; however, it would make distinguishing the different types of vegetation by color alone challenging. Although vegetation types would be difficult to distinguish, vegetation colors still remain distinct from the colors of the other worlds of the Solar System that do not show signs of surface life, as shown in Fig.7. The planetary data used for comparison to the Solar System planets with atmospheres, is based on data for Venus from Irvine (1968b), Mars and Jupiter from Irvine (1968a), and Saturn, Titan, Uranus and Neptune from Karkoschka (1998), following Hegde & Kaltenegger (2013).

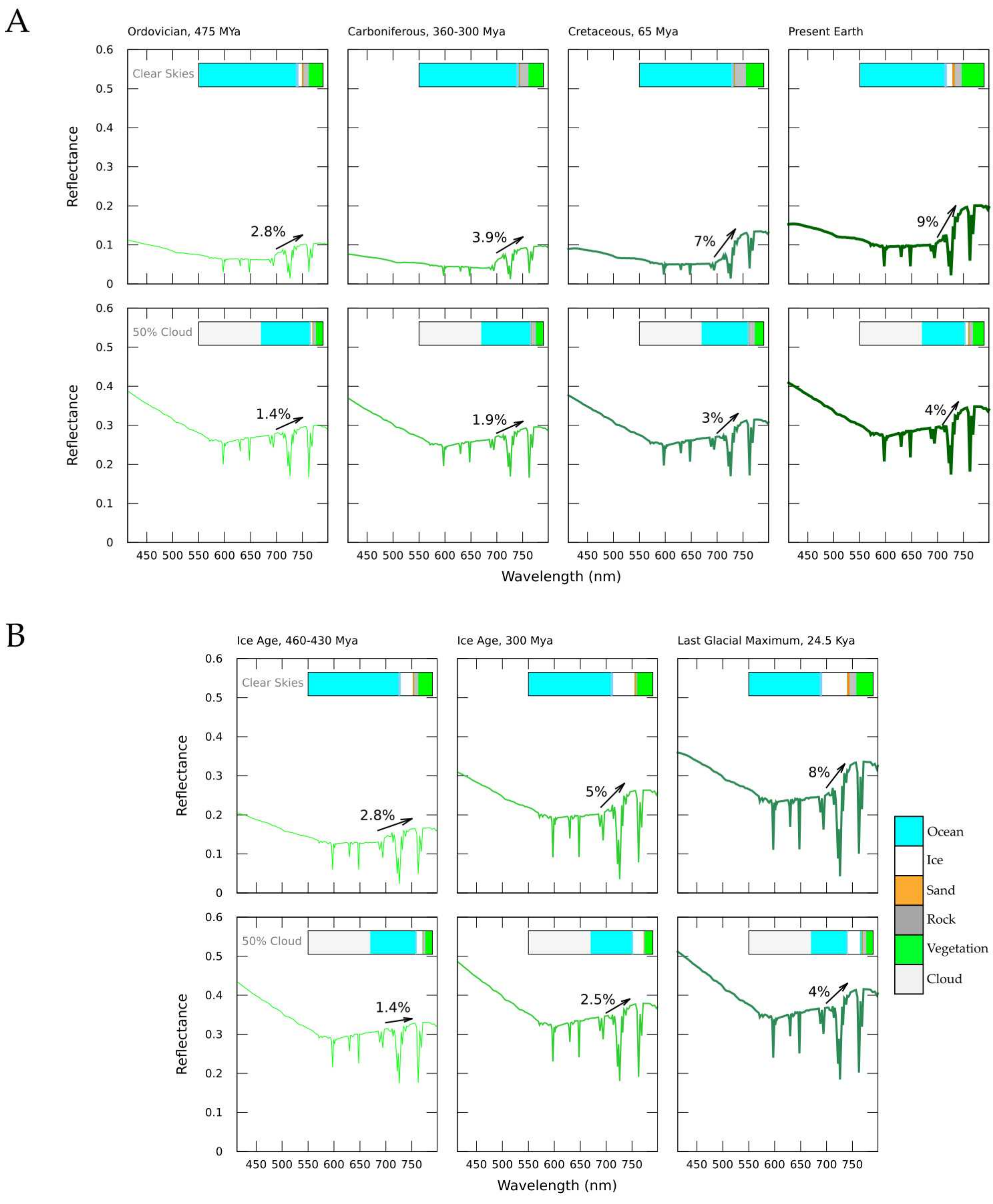

**FIG. 5.** Modeled VRE through Earth's geological time assuming equal land fractions of vegetation (60% on non-icy land surfaces) in each era, for (A) warm periods and (B) ice ages. Other surface features (ocean, land, ice, sand, rock) are varied for each era based on the values in Table 1.

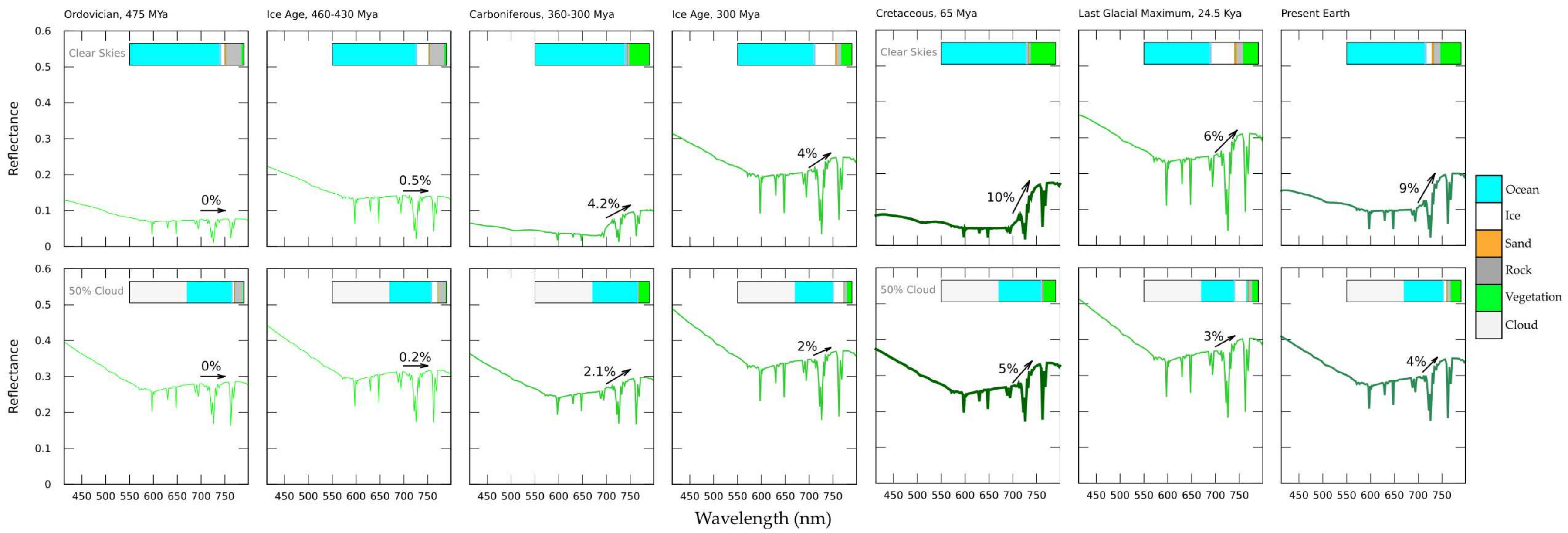

**FIG. 6.** Model VRE through geological time for our Earth (including effects of varied vegetation type and land, ocean and ice fraction).

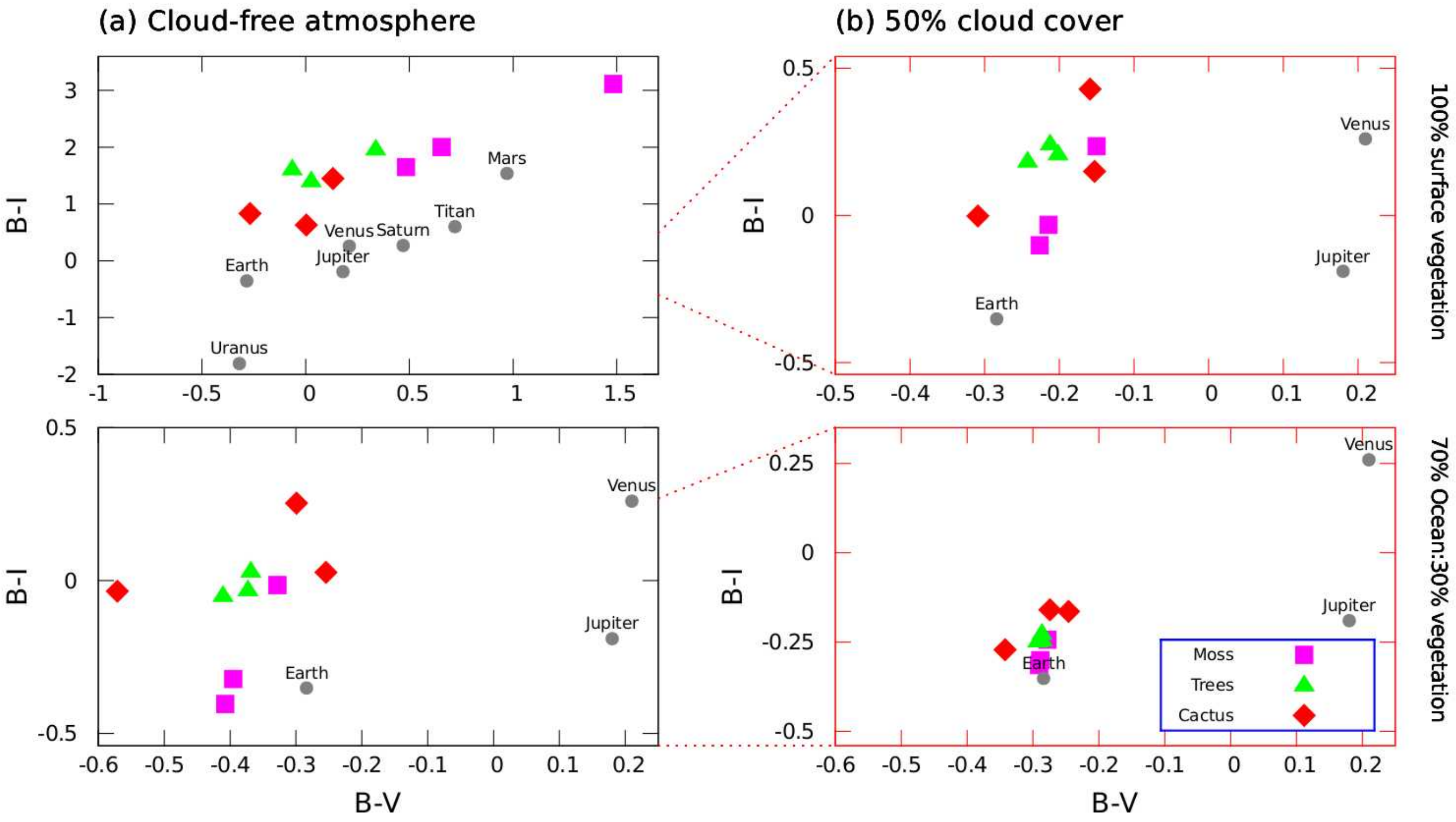

**FIG. 7.** Color-color diagrams for examples of vegetation at each of the three timeline epochs for a) clear atmosphere and b) present-day Earth-like cloud coverage of 50% (top panels) for 100% vegetation surface coverage and (bottom) and for a 30% continent (2% coast, 28% land):70% oceans distribution. We plot other solar system bodies for comparison (using spectra from Irvine et al. (1968a,b) and Karkoschka (1998)). Note the change of scale between panels.

## *4. Discussion*

### *4.1. Earth's evolutionary timeline is uncertain*

While fossil evidence and molecular clock analyses indicate that embryophytes (the higher plants outlined in Fig. 2) have been present on land for possibly 725 Myr (Zimmer et al. 2007), with fossil evidence suggesting < 500 Myr (Magallón et al. 2013), non-embryophytic photosynthesizers may have colonised the land as early as 0.85 to 1.2 Gyr ago (Horodyski & Knauth, 1994; Raven 1997; Knauth & Kennedy 2009). This hypothesis is based on chemical analyses of ancient rocks, so there is no direct fossil evidence from which to determine the characteristics of this early terrestrial life. However, early land ecosystems may have been composed of lichens, microbial mats or photosynthetic bacteria (Horodyski & Knauth, 1994; Raven 1997). Microbial mats would have incorporated photosynthetic microbes, producing an associated red edge reflectance feature (Sanroma et al. 2013). Lichens are formed from a symbiotic relationship between fungi and algae or cyanobacteria, which means they also have a red-edge feature that is comparable to terrestrial vegetation, especially mosses. Therefore, if ancient lichens were similar to modern equivalents, and if they were well-distributed on Earth's surface, the red edge signature could be much older than previously thought.

### *4.2. Identification of the VRE is challenging for exoplanets*

Our knowledge of the reflectivity of different surface components on Earth – like deserts, oceans and ice – helps in assigning the VRE of the Earthshine spectrum to terrestrial vegetation. Therefore the detection of the VRE on exoplanets, however interesting, will not be unambiguous. For example, some minerals can exhibit a similar spectral shape to vegetation around 750 nm (e.g. Seager et al. 2005). An open question is whether similar photosynthesis would evolve on a planet orbiting other host stars. If it does, the VRE could be shifted to different wavelengths (Kiang et al. 2007).

Only the overall VRE signal will be accessible for a disk integrated view of an exoplanet, therefore while the overall VRE strength on Earth increased with geological time, it will be challenging to distinguish different kinds of vegetation on an Earth-like planet, because unknown land:ocean fractions, overall vegetation surface areas, highly reflective ice fractions, as well as a range of possible cloud fractions will all influence the overall VRE signal.

### *4.3. Clouds decrease the detectability of vegetation and any surface features*

In our models we assumed an average cloud fraction of 50%, based on observed cloud cover on the present day Earth. However, changes in climate on Earth through geological time could have altered Earth's average cloud fraction. Whether this would have increased, or decreased, during warmer and colder periods is still open to debate. For example, warmer climates could increase humidity, favoring increased cloud formation (see e.g. Sellwood et al. 2000), but higher temperatures could also reduce nutrient cycling in the oceans, reducing the rate of biologically-produced cloud condensation nuclei, leading to optically thinner, shorter-lived clouds (Kump & Pollard 2008). Even if the overall cloud fraction remained approximately constant, cloud distributions could change (see e.g. Brierley et al. 2009, which shows how in a warmer climate high-cloud fractions could increase and low-cloud fractions could decrease).

### *4.4. Two possible Future Earth scenarios: Jungle Worlds or Desert Worlds increase the VRE signal*

As solar luminosity increases as a consequence of the Sun's main sequence evolution, Earth's surface habitats will be altered. Surface plant life is likely to become extinct as a result of the increased silicate weathering rate enhancing $CO_2$ drawdown, or the increased surface temperatures (O'Malley-James et al. 2013; 2014). However, before the end-point for surface vegetation is reached, the changes this gradual warming may have on the planet could result in very different habitats for future plant life. There are many uncertainties associated with how the climate system will respond to increasing solar luminosity and how life on Earth will adapt to meet the new challenges this creates. We cannot predict future vegetation on Earth; however, to explore the possible change of the VRE, we model two of many possible scenarios.

One potential future Earth scenario is a warmer Earth with more water vapor reaching the atmosphere, less ice cover and higher sea levels, resulting in a hot, humid planet similar to that of the Cretaceous era, which hosted dense global forests. Another scenario is that the future Earth becomes a hot, arid world (or transitions to hot, dry state after experiencing a hot, humid phase), if the extent of desert regions on Earth increases as the luminosity of the Sun increases. A rising mean surface temperature in equatorial regions could cause the extent of Hadley cells to increase in latitude, expanding desert regions (e.g. O'Malley-James et al. 2014), resulting in a hot, dry planet that favors desert plants over today's more common deciduous vegetation. Falling atmospheric $CO_2$ as a result of increased carbon draw-down (Lovelock & Whitfield 1982; Caldeira & Kasting 1992; O'Malley-James et al. 2013; 2014) could favor plants that are adapted to such conditions, like cacti, which use a carbon-concentrating photosynthesis method: Crassulacean acid metabolism (CAM; Keeley & Rundel 2003; O'Malley-James et al. 2014).

We model both possible scenarios: a hot, humid planet (a "jungle world") with deciduous forests and a hot, arid planet (a "desert world") using one form of desert-adapted vegetation, cacti, to explore the strength of a possible VRE signal on an older Earth (Fig.8), as well as on jungle and desert exoplanets in general. Note that there is a wide range of heat and drought tolerant vegetation species that could be dominant on a desert world, we chose one example, cacti.

For the jungle planet we assume that forests cover all land surface. For the desert planet shown in Figure 8 we assume 50% of the surface is covered by vegetation and 50% is bare sand. This assumption would require dense cacti vegetation in desert habitats on Earth.

For a clear atmosphere, a jungle planet (Fig. 8b, top) shows a VRE signal of up to 45%. Desert plants show a high NIR reflectivity, which is a consequence of adaptations to limit water-loss (Benson 1982; Nobel et al. 1986; Loik 2008 and references therein), causing them to reflect up to a factor of 2 more radiation at 750 nm than non-desert plants (Gates et al. 1965; Arnold 2008). However the high reflectivity of sand on a desert world (Fig. 8a, top) results in a weaker disk-integrated VRE signal of up to 30% in our model. Both scenarios would increase the strength of the VRE signal for a possible future Earth.

Arid planets should have a lower ocean fraction compared to Earth, while jungle planets should show higher ocean fractions. We compare the effect different ocean fractions would have on both types of planet in Fig. 8 (top panels). To show the effect of clouds (Fig. 8 (lower panels)) we set the desert planet ocean fraction to 60% to account for a warmer, drier climate than the present Earth. In addition we assume an increased ocean fraction of 80% for the jungle planet to account for a warmer, wetter climate than the present Earth. A humid jungle planet is more likely to have a higher cloud fraction than a dry desert planet; however, that cloud fraction could be lower than the present day Earth's (see Discussion). For a cloud fraction of 40% our model desert world VRE signal is about 11%, whereas our jungle planet shows a 7% VRE signal; both higher than the 4% VRE we model for the cloud-covered present day Earth.

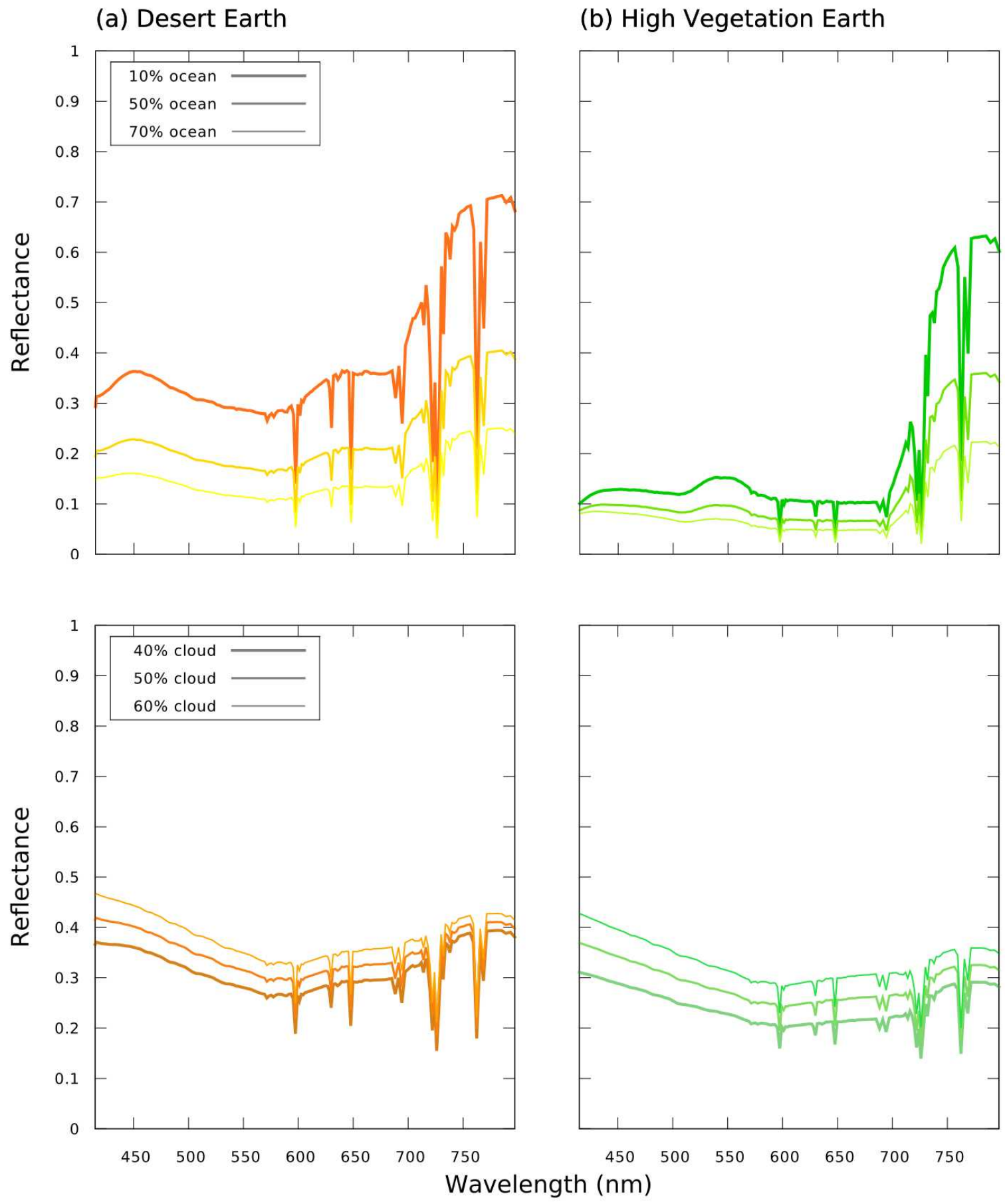

**FIG. 8.** Model VRE spectra of a desert world with cacti vegetation (a) and a jungle world with deciduous vegetation (b). (Top) a cloud-free Earth-like atmosphere with 10%, 50% and 70% ocean fractions. (Bottom) 40%, 50% and 60% cloud coverage.

Therefore, if we were to extend a possible VRE timeline into Earth's future, as it nears the end of its habitable lifetime we could see a slight increase in VRE strength for both scenarios if the planet enters a hot, humid stage (tempered by higher cloud cover) and/or a larger increase if the dominant climate state becomes hot and dry (aided by a lower cloud coverage).

### *4.5. Desert-worlds could be abundant*

Generally, habitable dry exoplanets (desert worlds with limited surface water) could be more common than water-rich planets like Earth, because the habitable zone for dry planets is wider than for Earth-like planets (Abe et al. 2011). Low levels of atmospheric water vapor allow habitable temperatures to persist at closer distances to the host star without causing a runaway greenhouse, while further out from the star, the lower levels of water allow the planet to resist freezing (Abe et al. 2011). The lack of water on desert worlds would also reduce cloud-cover, making surface features like the VRE easier to detect. If desert worlds have surface vegetation that has evolved a high reflectivity to conserve water loss, such planets provide interesting targets for successfully detecting a VRE signal. Such desert exoplanets with cactus-like vegetation and low cloud fractions would exhibit a strong VRE. Conversely, fast growing plants, which can complete their life cycles within a few weeks after a rare rainfall could be the dominant form of vegetation on desert worlds. This type of vegetation would lead to a short-lived, temporary VRE signal, which would be challenging to observe at the right time.

## *5. Conclusion*

If the development of life on other inhabited worlds follows a similar timeline to life on Earth, VRE biosignatures are only present for a fraction of a planet's habitable lifetime once vegetation covers the surface (from about 500 Myr ago on Earth). A VRE biosignature similar to that on the present-day

Earth would be difficult to detect without very high-precision instruments (e.g. Seager et al. 2005; Montañés-Rodríguez et al. 2006). However our models show that the VRE signature increases with geological time, as well as with increasing vegetation surface fraction and decreasing cloud coverage.

On Earth the VRE signature would have been weaker when the first land plants emerged about 500 Myr ago, like mosses, liverworts and hornworts as a result of the spectral properties of the first land plants. Once vascular plants, like ferns, emerged, plants had similar VRE strengths; however, the overall VRE signal increased due to globally increasing vegetation surface coverage. On Earth other geological surface feature changes, such as ice cover, modified the increasing VRE strength over time. The VRE signature generally increases through geological time, making older Earth-like planets better targets to find a vegetation surface feature.

The far-future evolution of Earth's climate is extremely hard to model. Two of the many possible scenarios, a warm humid future Earth (or jungle world) and a hot arid future Earth (or desert world), show an increased VRE signal in our models.

Generally, habitable dry exoplanets (desert worlds with limited surface water) could be more common than water-rich planets like Earth, because the habitable zone for dry planets is wider than for Earth-like planets. The evolution of high NIR reflectivity in desert vegetation on Earth suggests that exoplanets with hot desert climates could evolve highly infrared-reflective vegetation, producing a stronger VRE signature than Earth, assuming similar vegetation coverage.

Lower cloud coverage makes surface features easier to detect, leading to an increase in the detectable VRE signal. Therefore the best exoplanet targets for VRE detections should be older Earth-analogs, hot jungle worlds, and potentially, hot arid desert worlds, as well as exoplanets with low cloud coverage.

### *Acknowledgements*

The authors would like to thank an anonymous referee for insightful comments on an earlier version of this manuscript. We acknowledge the USGS Digital Spectral Library, ASTER Spectral Library and the Joint Fire Science Program spectral library. We also acknowledge funding from the Simons Foundation (290357, Kaltenegger).

### *Author Disclosure Statement*

No competing financial interests exist.

Address correspondence to:

*Jack O'Malley-James*
*Carl Sagan Institute,*
*Cornell University,*
*Ithaca, NY 14853, USA*

*Email*: jomalleyjames@astro.cornell.edu